\documentclass[12pt,preprint]{aastex}
\newcommand{\etal}{et~al.}
\newcommand{\xmips}{$24$~\micron$/2$-$8$~keV}

\newcommand{\hard}{$2$--$8$~keV}
\newcommand{\soft}{$0.5$--$2$~keV}
\newcommand{\cgsflux}{erg~s$^{-1}$~cm$^{-2}$}
\newcommand{\cmsq}{\hbox{cm$^{-2}$}}

\slugcomment{Paper status:  Accepted to the ApJS Spitzer Special Issue}

\shorttitle{X-ray--selected AGN at $24$~\micron}
\shortauthors{Rigby et al.}

\begin{document}

\title{$24$~\micron\ properties of X-ray selected AGN}
\author{J. R. Rigby\altaffilmark{1}, G. H. Rieke\altaffilmark{1}, 
R. Maiolino\altaffilmark{2}, R. Gilli\altaffilmark{2},
C. Papovich\altaffilmark{1}, P. G. P\'{e}rez-Gonz\'{a}lez\altaffilmark{1},
%late authors after this
A. Alonso-Herrero\altaffilmark{1}, E. Le Floc'h\altaffilmark{1},
C. W. Engelbracht\altaffilmark{1}, K. Gordon\altaffilmark{1}, 
D. C. Hines\altaffilmark{13}, 
J. L. Hinz\altaffilmark{1}, J. E. Morrison\altaffilmark{1}, 
J. Muzerolle\altaffilmark{1}, M. J. Rieke\altaffilmark{1},
\& K. Y. L. Su\altaffilmark{1}}

\altaffiltext{1}{Steward Observatory, University of Arizona, 933 N. Cherry Ave., Tucson, AZ 85721}

\altaffiltext{2}{INAF-Osservatorio Astrofisico di Arcetri, Largo Fermi 5, 50125 Florence, Italy}

\altaffiltext{3}{Space Science Institute, 4750 Walnut Street, Suite 205, Boulder, Colorado 80301}

\email{jrigby@as.arizona.edu}

\begin{abstract}
We examine the $24$~\micron\ to X-ray color of $157$ X-ray--selected AGN 
as a function of X-ray obscuration and optical classification in the Chandra Deep Field South.
The sample consists of the Chandra hard--band detections with 
\hard\ flux $>10^{-15}$~\cgsflux.
A deep $24$~\micron\ mosaic obtained with \emph{Spitzer} provides mid--infrared 
fluxes for the sample.
Since obscured AGN locally have higher $24$~\micron/\hard\ ratios than 
unobscured AGN, and since X-ray background models
predict a large population of obscured AGN, we expect
to find many X-ray--hard, IR--bright AGN.  Instead, we find that
the $24$~\micron\ to X-ray flux ratio does not depend on X-ray 
hardness in the full sample, nor does it differ between narrow--line 
and broad--line AGN.  We identify $5$ nearly Compton--thick AGN, and 
find they have similar $24$~\micron\ to X-ray flux ratios 
compared to the full sample.  
We consider AGN in the narrow redshift spikes at $z\sim0.7$;
for these AGN, there is some evidence that the flux ratio increases with 
X-ray hardness.  The redshift slice also shows an odd trend, which is also 
prominent in the full sample:
a group of X-ray--hard AGN with very low $24$~\micron\ to X-ray flux ratios.
\end{abstract}


\section{Introduction}
The hard spectrum of the X-ray background, which peaks at 
$30$--$40$~keV, implies the existence of a large population of 
obscured active galactic nuclei (AGN) \citep{madau,comastri,gillietal}.
Finding and understanding these obscured AGN is crucial to understanding 
the accretion history of the universe.  These AGN are predicted to be
particularly bright in the mid--infrared, as soft X-ray and ultraviolet
emission is absorbed by the intervening material and re-emitted by dust.
With \emph{ISO}, \citet{fadda} reported elevated mid--infrared emission
from obscured AGN, and \citet{frances} found the majority of 
\emph{ISO}--detected AGN to be obscured.

\emph{Spitzer} can test these ideas about the obscuration
of AGN and the interplay of the infrared and X-ray backgrounds.  
In this letter, by studying the Chandra Deep Field South (CDFS),
we present a first look at the \emph{Spitzer} mid--infrared properties of AGN.

\section{The Data and Catalogs}

We present deep new observations of the CDFS at $24$~\micron\ obtained
with the Multiband Imaging Photometer for Spitzer (MIPS)
\citep{rieke}.  These data were obtained with guaranteed time under
program 81; the relevant AORs are 8950528, 8951296, 8954112,
8957696, 8958208, 8958464, 8958976, 8959488, 8960000, 8960512,
8961024, and 8961536.  Slow scan--map mode was used, achieving
an average integration of $1380$~s.  The data were 
reduced using the DAT package of the MIPS instrument team \citep{gordon}.
The resulting mosaic 
%is $1.42^{\circ}$ by $0.49^{\circ}$ in size, and 
completely overlaps the $\sim17$\arcmin\ by $17$\arcmin\ Chandra field.  
The $80\%$ completeness limit, determined by adding artificial sources,
is $0.083$~mJy \citep{papovich}.  
This flux limit is comparable to the $5\sigma$ noise, which, scaling
from \citet{dole}, is $0.086$~mJy at the CDFS exposure depth 
(accounting for detector and confusion noise \citep{dole}.)
Thus, the full mosaic is one of the deepest, widest--field $24$~\micron\ 
maps yet obtained.

Sources were extracted from the $24$~\micron\ mosaic as 
described by \citet{papovich}.
In the $270$~arcmin$^2$ region having X-ray exposure times exceeding 
$0.5$~megaseconds, we detect $1147$ $24$~\micron\ sources brighter 
than the $80\%$ completeness limit.
In the $200$~arcmin$^2$ region having $>0.75$~Ms X-ray coverage, we detect
$878$ such $24$~\micron\ sources.
(The X-ray depth of coverage varies across the field because the $11$
observations comprising the full Chandra exposure have different roll angles.)

% MOVE THIS TO BEST SPOT!%%%% NEW HERE!
Two papers currently in preparation will publish $24$~\micron\ fluxes 
for $\sim90$ X-ray sources in CDFS \citep{rigbya,rigbyb}; most
of these sources are within the main sample of this letter.

The Chandra 1~megasecond CDFS is one of the deepest X-ray observations
to date.  Two reductions and source catalogs exist in the literature.  
The first is by the CDFS team, who
find $304$ sources detected jointly by SExtractor and wavdetect 
%as well as $28$ SExtractor--only and $14$ wavdetect--only sources 
(Giacconi \etal\ 2002, hereafter G02).
\nocite{giacconi}
The second is the CDF North team's reduction, using
the same processing and source extraction as for the 
North (Alexander \etal\ 2003, hereafter A03).  \nocite{alex}
Their source catalog has $326$ sources.  A03 compare these 
two catalogs, finding $293$ sources in common.
%; sources detected in one catalog but not the other are generally of lower
%significance or at large off--axis angles.
Because the \emph{Spitzer} cosmological surveys will also target the CDF North, 
in this paper we use the A03 catalog, to facilitate future
comparisons between the North and South fields.

The astrometry of the CDFS Chandra catalogs is excellent.
A03 registered their astrometry to the frame of the R-band ESO Imaging Survey
\citep{arnouts}; the median separation between X-ray and 
R--band positions is $0.37$\arcsec\ (A03).  Using the same R--band 
catalog,\footnote{From CDS, at http://vizier.u-strasbg.fr}
we registered our $24$~\micron\ mosaic to their astrometric frame 
by adding a $0.52$\arcsec\ global shift.

The other catalog used in this paper is that of Szokoly \etal\ 2004 
(hereafter S04), who present secure spectroscopic
redshifts for $137$ X-ray--selected sources in the CDFS.  They
classify their sources in two ways:  based on optical spectra, and
based on X-ray luminosity and hardness.
\nocite{szokoly}

\section{Sample Selection}

We select hard band (\hard) detections in the CDFS from the catalog
of A03; there are $198$.  Selecting hard band detections (rather than
soft or any-band detections) reduces contamination from non--active
galaxies.  To further maximize the AGN fraction, we
require the hard--band flux to exceed $10^{-15}$~\cgsflux, 
which reduces the sample to $157$ sources.  
This is our main sample.  None of these sources 
is classified by S04 as a star or a non--active galaxy.
Almost all of these sources are likely to be AGN (see Figure~2 of \citet{barger});
65 have been identified as AGN by their X-ray hardness and luminosity (which
requires a known redshift), 
and 39 have been identified as AGN by optical spectroscopy.

Because the soft band of Chandra (\soft) is more sensitive
than the hard band, sources in our main sample generally have a
well--measured soft-band flux or a well--constrained limit.

For the main sample, we seek redshifts as available from S04.  S04
targeted optical counterparts to X-ray sources from the G02
catalog (and thus associate redshifts with G02 X-ray sources),
whereas we use the A03 X-ray catalog.  As a result, it is
necessary to associate the A03 and S04 catalogs, which we do by
matching sources whose coordinates are separated by less than $2$\arcsec.
Such a large separation is necessary to match sources at large off--axis angles,
where Chandra centroiding is difficult.  Since the only spectra in the
S04 catalog are counterparts to X-ray sources, the source density is small
($\sim 0.5$ arcmin$^{-2}$), so the probability of false matches with the A03
X-ray catalog is small.
In the main sample, $65$ AGN have secure redshifts and optical classifications.

For each X-ray source in our main sample, we search for a
$24$~\micron\ counterpart from our catalog.  To claim detection, we
require the fractional flux uncertainty to be $<50\%$ and the
coordinates (X-ray and $24$~\micron) to be offset by $<1.2$\arcsec.
(This small separation is sufficient to match sources over most of the field,
and is necessary to prevent false matches, since the $24$~\micron\ source
density is high.)  No X-ray source has multiple $24$~\micron\ sources within
the permitted offset.
Given the $24$~\micron\ source density reported above, for the main
sample of $157$ X-ray sources, by chance we expect $\sim1$ false $24$~\micron\
counterpart using the P-statistic (see e.g. \citet{lilly}).
The result of the counterpart search is that $98$ of the $157$ sources
in the main sample have $24$~\micron\ counterparts.  For
sources not detected at $24$~\micron, we take the flux upper limit to
be the $80\%$ completeness limit ($0.083$~mJy). 

\begin{figure*}
\figurenum{1}
\includegraphics[width=18cm]{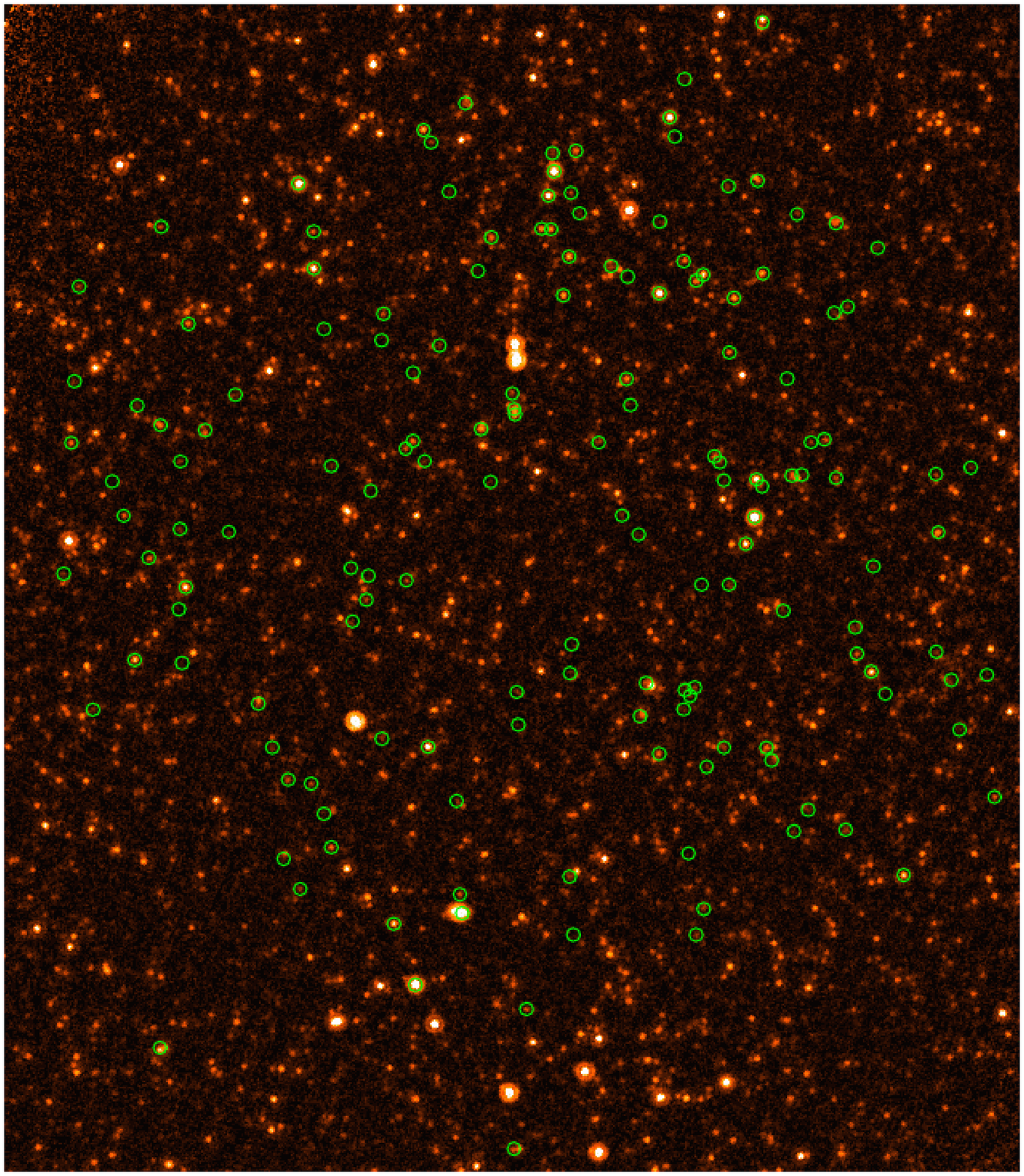}
\figcaption{$24$~\micron\ image of CDF-S.  Overplotted are the X-ray positions of
the main sample.  The circles are large (R$=8$\arcsec) for presentation purposes.
The scale is $20.3$\arcmin\ by $23.3$\arcmin; North is up and East is left.
\emph{Figure has been compressed (and thus degraded) for preprint 
distribution; please see the Journal for full quality figure.}
}
\label{fig:plate}
\end{figure*}

Figure~\ref{fig:plate} displays the section of the $24$~\micron\ mosaic that 
overlaps the Chandra field, with the main sample X-ray positions overlaid.

\section{Comparing X-ray and $24$~\micron\ colors}

AGN are traditionally classified by optical spectroscopy.  However,
for distant sources, nuclear emission can be overwhelmed by stellar
emission, especially for low--luminosity AGN \citep{moran}.
Obscuration or lack of the narrow lines may also play a role \citep{maiolino}.
Consequently, $43\%$ of X-ray selected AGN in the CDFS are not classified
as active by optical spectroscopy (S04).
Because of this, many authors have argued that X-rays are a better way to 
\emph{select} AGN than optical spectroscopic surveys.  

Further, X-rays may be a more effective way to \emph{classify} AGN \citep{hasinger}.  
AGN with harder X-ray spectra are generally
measured to have thicker columns of absorbing gas than do X-ray--soft AGN.
Similarly, in unification models, optical Type~2 AGN are drawn
from the same population of objects as Type~1 AGN, but they are viewed
through thicker columns of obscuring gas and dust.  Thus, X-ray
hardness can estimate the obscuration of AGN when optical spectroscopy
cannot.

\begin{figure}
\figurenum{2}
\includegraphics[angle=270,width=8cm]{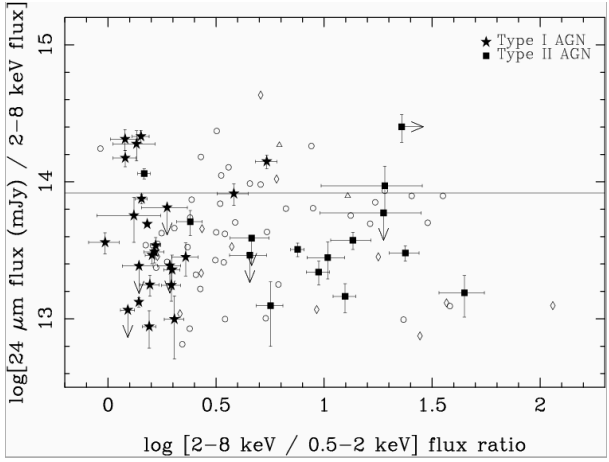}
\figcaption{Comparison of optical and X-ray AGN classification.
Symbol types indicate classification from optical spectroscopy:
broad--line AGN and QSOs (\emph{star symbols}) as well as AGN with
high--excitation narrow lines (\emph{filled squares}).
These are the classical Type~1 and Type~2 categories, respectively.
%, and are well--separated
%by the X-ray flux ratio; the Type~1 AGN cluster at soft X-ray flux ratios.
Also plotted are sources whose optical spectra show
low--excitation emission lines (\emph{diamonds}) or
absorption lines (\emph{triangles}), as well as sources lacking 
optical classification (\emph{circles}).
To improve plot clarity, for the last three categories, sources with
non-detections in $24$~\micron\ or soft X-ray are omitted, and errorbars are omitted.
Optical spectroscopy and classification are from S04.
}
\label{fig:optxray}
\end{figure}

Figure~\ref{fig:optxray} compares optical and X-ray classification of
the CDFS AGN.
The x-axis is an indicator of spectral hardness: the
ratio of the \hard\ flux to the \soft\ flux, which we define as 
H/S $=$ $\log$[flux(\hard)/flux(\soft)].
Sources with broad emission lines (optically--selected Type~1 Seyferts
and QSOs) cluster toward the X-ray--soft side of the figure.
These objects are well--separated in X-ray hardness from the sources
with optical high--excitation narrow emission lines
(optically--selected Type~2 AGN), which span a range of higher X-ray
hardness.  This clear separation validates the use of the hard/soft
X-ray flux ratio H/S as a tool to classify AGN, as also found by
\citet{mainieri}.  Indeed, S04 also used X-ray hardness to classify
these CDFS AGN, using the same Chandra data and very similar energy
bands to this work, but a counts ratio rather than a flux ratio; their
division between Type~1 and Type~2 AGN lies at H/S$\approx0.6$ in our
Figure~\ref{fig:optxray}, and agrees well with the boundary between
optically classified Type~1 and 2 AGN.

In Figure~\ref{fig:24hs}, we now examine how H/S affects the behavior
of the $24$~\micron\ to X-ray flux ratio, defined as 
24/X $=\log$[($24$\micron\ flux density, mJy)/(\hard\ flux, \cgsflux)].
There is a factor of $100$ spread in both the H/S and 24/X flux ratios.
The spread in H/S reflects the range of X-ray hardness found in AGN.  
The reasons for the  spread in the 24/X flux ratio we discuss below.

The horizontal line in Figures~\ref{fig:optxray} and~\ref{fig:24hs}
is the 24/X color of a source with hard X-ray flux at the sample
cutoff, and $24$~\micron\ flux at the $80\%$ completeness limit.
Since the sample is X-ray selected, we are reasonably complete to colors
above the line, while we may be missing sources with colors below the 
line (because the $24$~\micron\ flux may fall below the detection limit.)

Figures~\ref{fig:optxray} and~\ref{fig:24hs} show that overall,
the \xmips\ flux ratio does {\bf not} depend on the hardness of the AGN.
There is no global trend of the 24/X flux ratio with H/S.
Nor does the 24/X ratio differ significantly 
between the X-ray--classified hard and soft AGN in Figure~\ref{fig:24hs},
nor between the optically--classified Type~1 and Type~2 AGN in 
Figure~\ref{fig:optxray}.  
 
\citet{fadda} found that $5$--$10$~keV XMM sources were twice as likely 
to have $15$~\micron\ counterparts compared with \soft\ sources;
however, this result was based on just $28$ IR--detected hard X-ray sources.
The authors used the IR/X-ray spectral slope to classify sources as 
Type~1 or Type~2 AGN or dominated by star formation, and then applied these 
classifications to estimate the AGN contribution to the IR background.
(They did find some X-ray--soft AGN with high IR/X ratios, which they attributed 
to host galaxy contamination.)
As discussed above, our full sample (Figure~\ref{fig:24hs}) contradicts
the  \citet{fadda} result that the mid--IR flux of AGN 
is a strong function of hardness.

\section{Discussion}

To understand the behavior of the 24/X flux ratio, in
Figure~\ref{fig:templates} we present the flux ratios predicted by
local AGN templates with a range of obscuration \citep{silva}.  These
templates were obtained by interpolating nuclear infrared photometric
data of a sample of 33 Seyferts.  Each infrared SED was normalized to
the absorption--corrected hard X-ray luminosity, and then averaged
within bins of absorbing N$_H$.  Using these empirical templates,
\citet{silva} derived the contribution of AGN to the mid-IR
background and successfully matched the observed fraction.

In Figure~\ref{fig:templates}, at $z=0$ as the obscuring column
increases, the predicted 24/X ratio increases (since the $24$~\micron\
flux rises due to reprocessing and the
\hard\ flux falls dramatically due to absorption.) This effect grows weaker with 
increasing redshift because of K-corrections:  as the \hard\ band samples
progressively higher rest--frame energies, it is less absorbed for a 
given column.  
Generally in the deep X-ray surveys, Type~1 AGN have $\log$N$_H \la 22$~\cmsq,
whereas Type~2 have $22\la\log$N$_H\la24$~\cmsq\ \citep{gilli}.

\begin{figure}
\figurenum{3}
\includegraphics[angle=270,width=8cm]{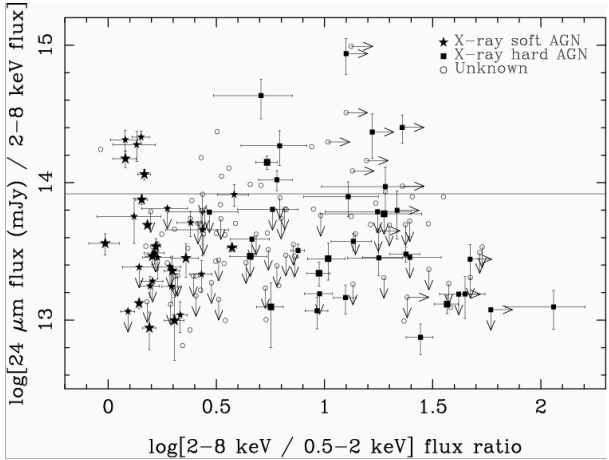}
\figcaption{IR/X-ray color versus X-ray hardness.
All sources in the main sample are plotted, and
symbol types indicate X-ray classification from S04:
X-ray--classified soft AGN (\emph{stars});
X-ray--classified hard AGN (\emph{solid squares});
and unclassified sources (\emph{circles}).
Sources lacking measured redshifts are ``unclassified'', since
a redshift is necessary to measure the X-ray luminosity, which S04 
required along with X-ray hardness to classify sources.
For clarity, errorbars are not plotted for the unclassified sources.
Larger symbols indicate QSO luminosities.
}
\label{fig:24hs}
\end{figure}

\begin{figure}
\figurenum{4}
\includegraphics[angle=270,width=10cm]{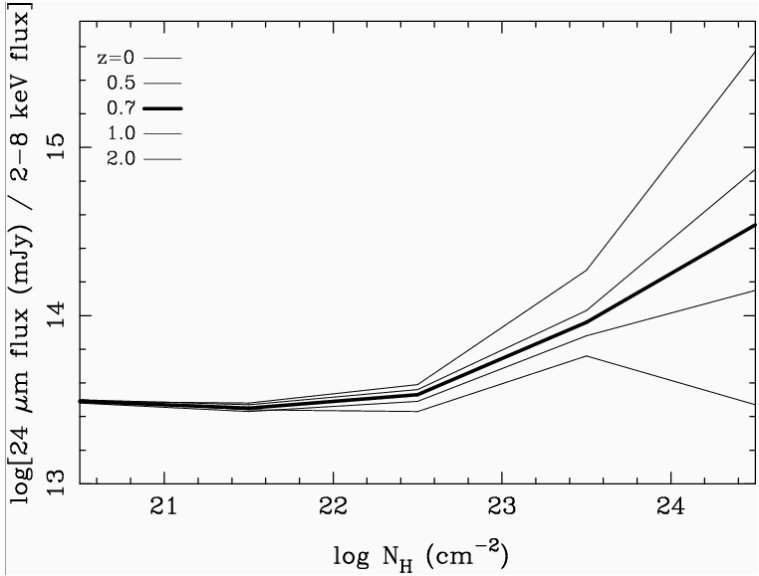}
\figcaption{Behavior of \xmips\ flux ratio with increasing obscuration.
From top to bottom, lines show line ratios for local templates \citep{silva}
redshifted to $z=0$, $0.5$, $0.7$ (\emph{dark line}), $1.0$, and $2.0$.  
Intrinsic differences in the template SEDs create a 24/X scatter of about 
$\pm0.5$~dex around the mean values plotted.
}
\label{fig:templates}
\end{figure}

A basic conclusion from Figure~\ref{fig:templates} is that local AGN 
with $\log N(H) \la 23.2$~\cmsq\ cannot produce 24/X $>14$ for $0<z<2$.  
When X-ray--soft AGN are observed to have such high 24/X ratios (as seen 
for a minority of sources in Figure~\ref{fig:24hs}),
this suggests that the $24$~\micron\ flux is dominated by 
star formation in the host galaxy, not by accretion.  (Some of
the harder sources with high 24/X may also be star--formation dominated.)
Using this rule of thumb, we see that star formation does not dominate most 
of the sources in Figure~\ref{fig:24hs}; their 24/X ratios are consistent 
with AGN power.  (See also \citet{almudena}.)

\begin{figure}
\figurenum{5}
\includegraphics[angle=270,width=10cm]{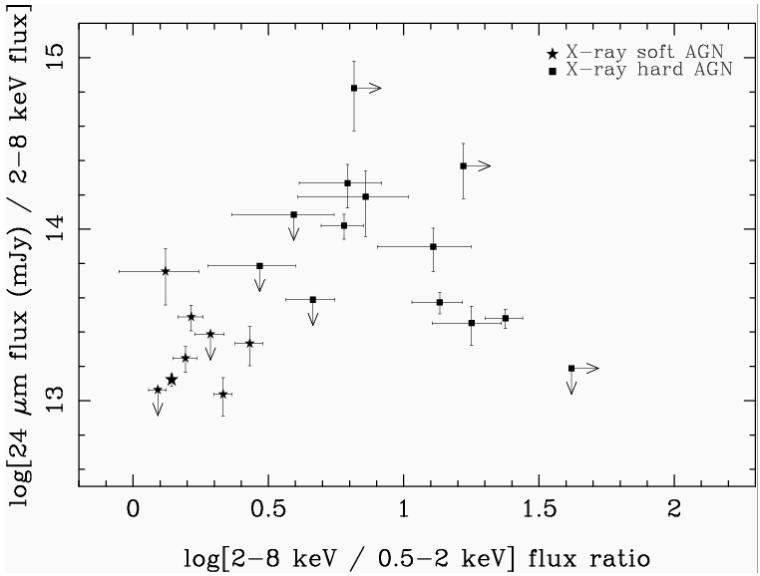}
\figcaption{IR/X color for the $0.65<z<0.75$ AGN of the CDFS.  All AGN with spectroscopic
redshifts in this range have been plotted; no X-ray flux cut has been imposed.  
Symbols and classification as in figure~\ref{fig:24hs}.
}
\label{fig:z7}
\end{figure}

To disentangle the effects of redshift and compare the distant AGN to
local templates,
in Figure~\ref{fig:z7} we plot all hard--band detected AGN in the 
redshift range $0.65<z<0.75$.
This range encompasses the prominent $z=0.674$ and $z=0.734$ 
redshift spikes (S04), which allows us to restrict the redshift range 
(and thus the spread in K-correction) while maximizing the sample size.

We first examine the X-ray--soft AGN in Figure~\ref{fig:z7}.  
Their 24/X ratios agree well with the local templates, which predict 
24/X$\approx 13.5$ for AGN with $\log$ N$_{H}< 22.5$~\cmsq\ at this redshift.
The spread in template SEDs predicts a scatter in the flux ratio of 
about $\pm0.5$ in log units, which is consistent with the data.  

We now examine the 24/X ratio for the hard (H/S$\ga0.6$) AGN.
We propose two possible interpretations of the trends in Figure~\ref{fig:z7}:
\begin{itemize}

\item There is no trend of 24/X with X-ray hardness in Figure~\ref{fig:z7}.
The distribution simply reflects the scatter seen in Figure~\ref{fig:24hs}.  

\item The 24/X ratio does indeed increase with hardness in this 
subsample, as expected from reprocessing: H/S$\sim 0.8$ have higher 24/X than 
the H/S$<0.5$ AGN.  However, there are several strange exceptions:
AGN with very hard X-ray flux ratios and very low 24/X.

\end{itemize}

This population of X-ray--hard AGN with low 24/X is also seen in the 
full redshift sample (Figure~\ref{fig:24hs}), where there are $\sim15$ 
such sources.  Their 24/X ratios are $0.5$--$1$ dex lower than predicted 
by the templates.
These sources are surprising within the unification paradigm: the
hardness ratios indicate large column density, which unification 
scenarios associate with a dusty molecular torus; this column should absorb
the X-ray radiation, thus increasing the 24/X ratio relative to 
lower--column AGN.  Additionally, dust in the torus should absorb a large
fraction of the UV continuum and re-emit that energy in the
mid--infrared, further increasing 24/X.  In the X-ray--hard
$24$~\micron--faint sources in question, this expectation is not
observed.
Thus, these sources merit further investigation; 
there may be a real difference between these AGN and local AGN 
(for example in their dust properties or absorbing geometries.)
Though the average 24/X value for the X-ray--hard AGN may be higher than 
it appears in Figure~\ref{fig:24hs} due to selection effects, the sources 
with low 24/X still beg explanation.

While the H/S X-ray flux ratio is a useful diagnostic, the more fundamental  
parameter is the amount of obscuring column.  
For the very hard (H/S$>1$) AGN with known redshifts, we examine the X-ray 
spectral information derived by \citet{gilli}. 
Ten of the $16$ spectra are fit by 
$\log$N$_H \approx 23$~\cmsq, while only $5$ fits require $23.7<\log$N$_H<24$~\cmsq.
Those five nearly Compton--thick AGN have 24/X values  spanning 
$13.2$ to $14.4$, which is the same range populated by the other AGN (soft and hard).

Could we have missed the majority of the highly obscured, high 24/X AGN?
Since Compton--thick AGN should be fainter at \hard\ than less obscured AGN, 
it is quite possible that the most obscured AGN are the faintest in our survey.
The faint (hard flux $\la 3\times 10^{-15}$~\cgsflux) AGN are less likely to 
have known spectroscopic redshifts and reliable X-ray spectral fits.
Since they are faint in the hard band and undetected in the soft, they 
have lower limits on H/S that permit but do not require extreme hardness.
Several such objects have high 24/X.  Thus, these AGN are good candidates for being highly
obscured.  If they are, they would populate the high H/S, high 24/X region
of Figure~\ref{fig:24hs} where obscured AGN are expected, but not seen.

\section{Conclusions}
We present the first look at the $24$~\micron\ to X-ray colors of
X-ray--selected AGN.  The expectation that harder AGN should be relatively
brighter at $24$~\micron\  is not confirmed.  
Rather, there is large scatter and 
no trend in 24/X color with X-ray hardness in our sample of $157$ AGN 
(with dynamic range of 100 in both flux ratios).
In addition, the 24/X colors of optically-identified Type~1 and~2 AGN 
are indistinguishable.
%Based on 24/X color, star formation appears to dominate a small 
%fraction of $24$~\micron\ fluxes.

In a narrow redshift slice (picked to minimize differential K-corrections), 
we see hints that 24/X may increase with X-ray hardness (though this effect
is not seen in the full sample).  However, in both the restricted--redshift
and full samples, we identify AGN with unusual colors:  very hard X-ray flux ratios
and surprisingly low 24/X ratios.  These AGN merit further investigation.

X-ray spectral fits to $16$ of the X-ray--hard AGN find column densities of 
 $10^{23-24}$~\cmsq, most toward the lower end of this range.  Thus,
only a handful are nearly Compton--thick (N$_H\approx 10^{24}$).  Even
these sources do not distinguish themselves in 24/X color from the 
lower--column AGN.

Thus, while this initial survey has not yet confirmed expectations for a
population of X-ray--obscured, infrared--bright AGN, we have 
raised new questions:  What explains the factor of $\sim100$ range in
$24$~\micron\ to \hard\ flux ratio?  Why do many of the hardest AGN have 
very low 24/X ratios compared with local templates?  And if hard X-ray selection
and $24$~\micron\ follow--up finds few Compton--thick AGN, what is a better
way to search?
Future surveys and follow--up observations are clearly needed to 
address these questions.

\acknowledgements
We thank all who made \emph{Spitzer} a reality.
We also thank the CDF North team for making available Chandra fits images
of the CDFS.
This work is based in part on observations made with \emph{Spitzer}, 
which is operated by the Jet Propulsion Laboratory, 
California Institute of Technology under NASA contract 1407. Support for 
this work was provided by NASA through Contract Number 960785 issued by 
JPL/Caltech.

%%%%%%%%%%%%%%%%%%%%%%%%%%%%%%%%%%%%%%%%%%%%%%%%%%%%%%

%%%%%%%%%%%%%%%%%%%%%%%%%%%%%%%%%%%%%%%%%%%%%%%%%%%%%%%%%%%%%
\clearpage

\end{document}